\documentclass[referee]{aa}
\usepackage{mathptmx}
\usepackage[T1]{fontenc}
\usepackage{ae,aecompl}
\usepackage{graphicx}	
\usepackage{amsmath}	
\usepackage{amssymb}	
\begin{document}
\title{Equilibrium solution for cold dynamical systems and self-similarity.}
\author{Alard, C.}
\institute{Institut d'Astrophysique de Paris, 98bis Boulevard Arago, Paris}
\date{}
%
%
%
\abstract{}
{Numerical simulations demonstrate a link between dynamically cold initial solutions
and  self-similarity. However the nature of this link is not fully understood.
Cold initial conditions alone without further symmetry do not lead to self-similarity.
}
{Here we show that when the system approaches equilibrium a new symmetry appears. 
The combination of this equilibrium symmetry with the cold symmetry in the initial conditions
leads to full self-similarity. As a consequence for any initially cold system even if the initial spatial distribution
is not self-similar we will observe an evolution towards self-similarity near equilibrium.}
{The case of one dimensional systems
or spherically symmetric systems in 3D are discussed in detail. Systems depending on the energy and other integrals are also considered.
The problem of the degeneracy of the self-similar solutions at equilibrium is tackled. It is shown that very small perturbations at the center
of the system have the ability to break this degeneracy and lead to the convergence towards a specific auto-similar solution.}
{}
\maketitle
\section{Introduction}
\label{Intro}
 The cold dark matter (CDM) paradigm in cosmology implies that structures formed
 from initial conditions with a very small
 velocity dispersion. High resolution numerical simulations of CDM structure formation 
 show that a self similar regime (see \cite{Gunn}) appears in the central 
region of halos and extends for about two decades (see \cite{Ludlow}). The origin
 of this self similar regime and the convergence towards a self similar solution
 near equilibrium is well established for specific conjectures (see \cite{Bertschinger1985}  
\cite{Binney}, \cite{Lancellotti}, \cite{Alard2013b}, \cite{Halle},
 \cite{Schulz}). However a fundamental problem remains. How is it possible that 
starting from various non self-similar initial conditions the system evolves after
a number of dynamical times towards self-similarity ?
Furthermore since a large class of self-similar solutions exists, how does the system 
makes a choice and evolves towards a specific
similarity index ? Here are the questions we will try to answer in this work.
\section{The road to equilibrium for a cold system}
\begin{figure*}
 \includegraphics[width=17cm]{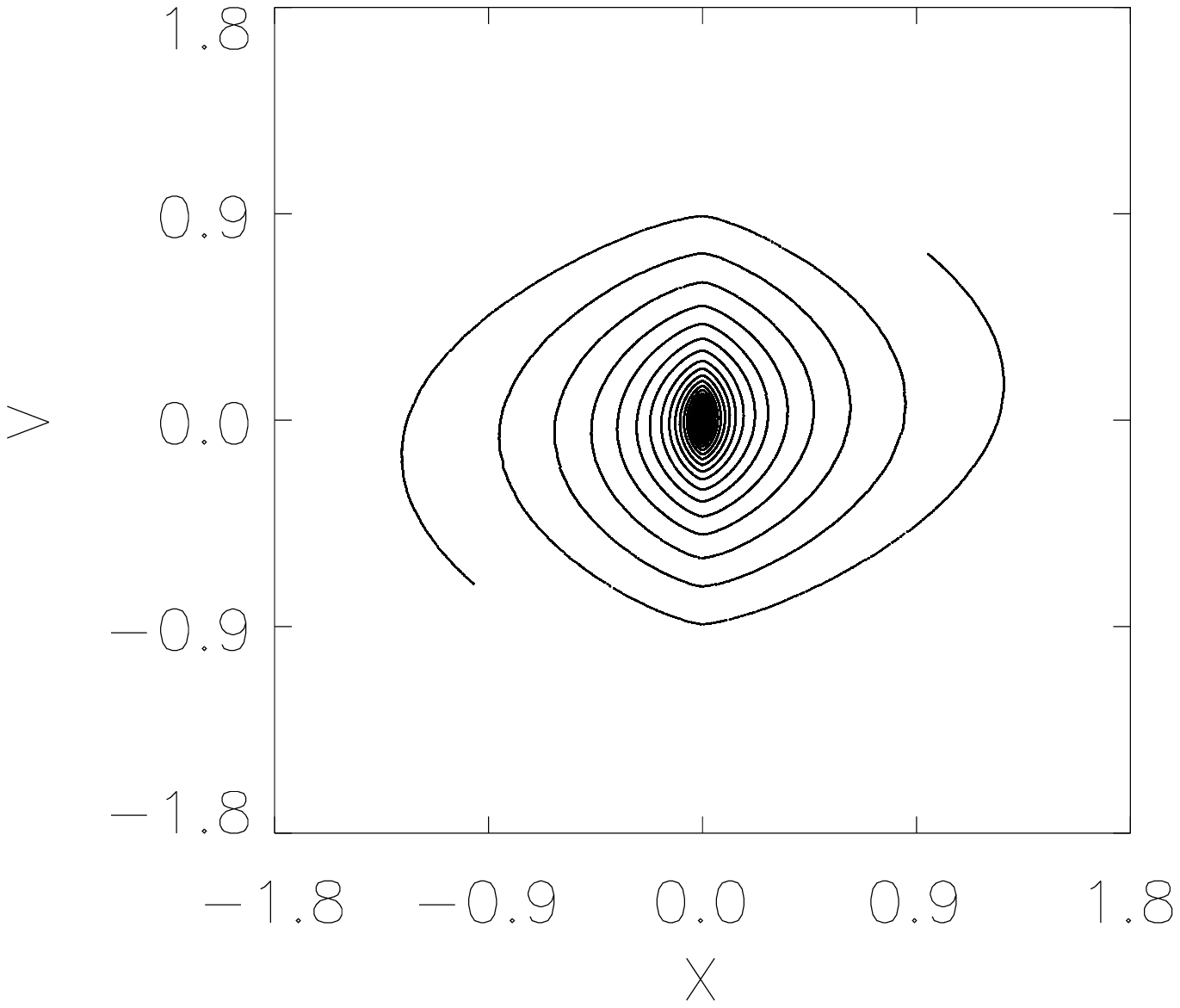}
  \caption{The late evolution of a system with cold initial conditions for 1D gravity. The density profile of the
  initial conditions in the $x$ dimension is a power-law with slope $\frac{-1}{2}$.}
 \label{Fig1}
\end{figure*}

\begin{figure*}
 \includegraphics[width=17cm]{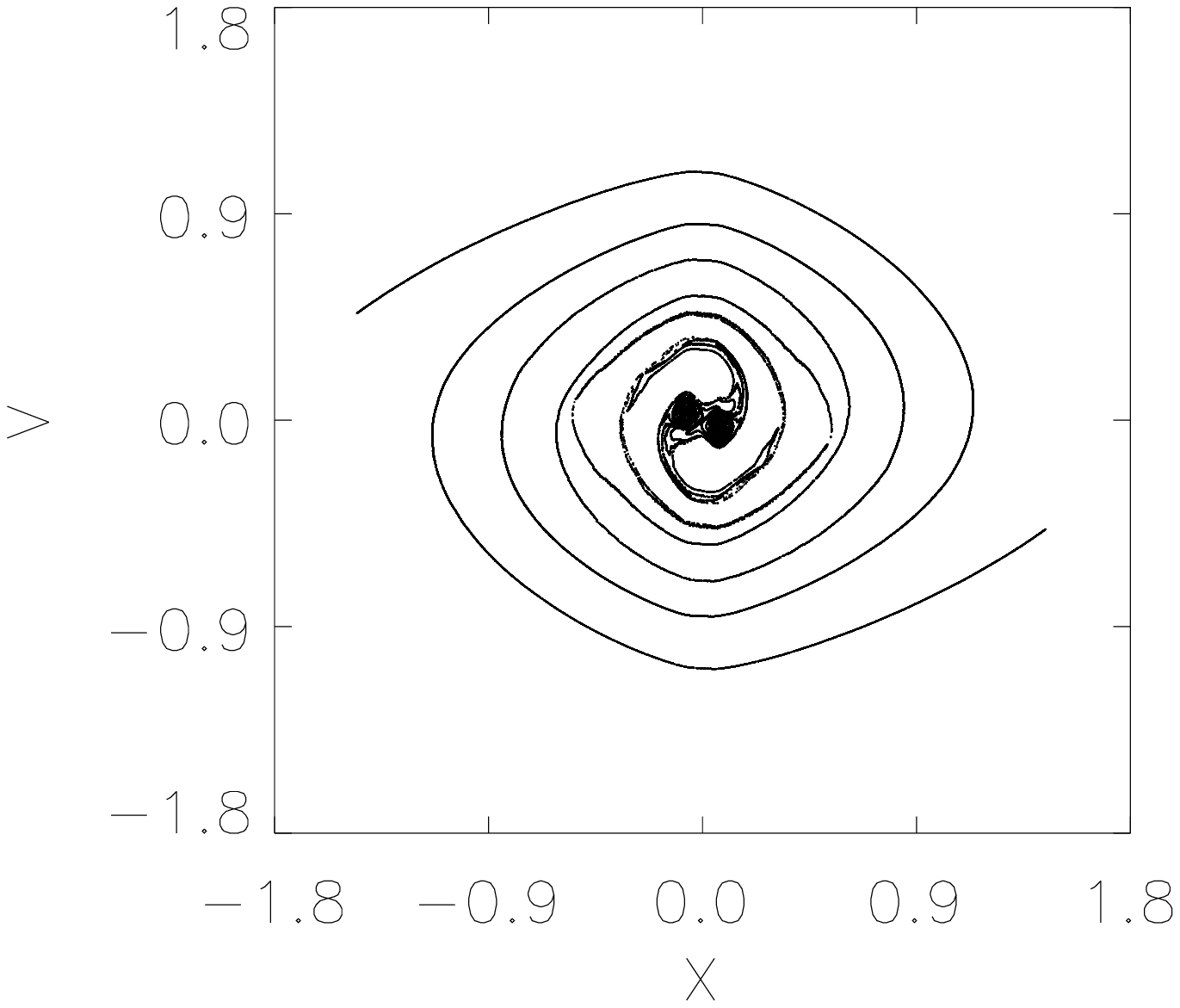}
  \caption{The late evolution of a system with cold initial conditions for 1D gravity. The density profile of the
  initial conditions in the $x$ dimension is two power-laws with slope $\frac{-1}{2}$ and different centers. Note that
  the structure of the self-similar solution of Fig (\ref{Fig1}) is completely lost. The dynamic in the central region is dominated
  by oscillations between the 2 power-law cores. The relaxation process and the strength of the gravity in 1D is not strong enough
  to overcome this oscillation mode.}
 \label{Fig2}
\end{figure*}

 As the system evolves from the initial conditions the number of orbits made by a given point increases.
 For each new orbits a fold develops and as a consequence the number of folds in phase space increases quickly.
 Increasing the number of folds has the obvious consequence to increase the elongation in phase space.
 If we consider the evolution of an initial small area in phase space as the system evolves this area become more and more
 elongated. Provided that the area is small enough the phase space density in the elongated area is nearly constant.
 When the elongation become large this small area has enough extension to cover a whole fold. As a consequence
 the density in phase space along the fold is nearly constant. Thus this fold represents an iso-contour of
 the density in phase space. Let now imagine that we smooth the solution with an adaptive filter. The local scale
 of this filter is of the order of the inter-fold distance. The smoothing operation using this filter would not produce
 a continuous density if the elongated fold with nearly constant density is not nearly closed. As a consequence with
 increasing elongation the fold must become nearly closed. A nearly closed fold is basically a fold near equilibrium.
 This reasoning indicates that after a sufficient number of orbits a cold system should reach a state which is close to equilibrium. 
 A process that could change the evolution of the system and could stop the increase in the number of folds
 and the increase of the local elongation would the appearance of global oscillations. For instance if an asymmetry
 is present at the center a stable oscillation may be produced which would make the convergence towards equilibrium
 impossible. An example of such process is illustrated In Fig.'s (\ref{Fig1}) and (\ref{Fig2}). The evolution of a system with two distinct cores
 in the initial conditions id presented in Fig. (\ref{Fig2}). In contrast to Fig. (\ref{Fig1}) where a single core is present the system in Fig. (\ref{Fig2})
 does not converge towards a self-similar solution. The situation may have been different in 3D where the gravity is stronger and thus where the relaxation
 process is more efficient in its ability to erase the details of the initial conditions.
\section{Equilibrium as a symmetry}
\label{Sec_Eq}
 At equilibrium various quantities like the energy or others integrals of motion are conserved along individual orbits. Thus the density
 at equilibrium is a function of the energy and of other the integrals of motion. 
\subsection{Multi-dimensional systems depending on a single spatial and velocity variable.}
In this section we will consider either one-dimensional gravity
 or 3 dimensional gravity with spherical symmetry. In such case the density in a phase 
 space with general coordinates $(x,v)$ can be expressed as a function of the energy 
 and other integrals like the angular momentum.
\subsubsection{Systems depending only on the energy.}
\label{sec_1}
\begin{equation}
 f(x,v) \equiv F(E) = F\left(\phi(x)+\frac{v^2}{2} \right)
 \label{eq1}
\end{equation}
For convenience we re-scale the velocity axis, then Eq. \ref{eq1} reads,
\begin{equation}
 f(x,v) \equiv F(E) = F \left(\phi(x)+v^2 \right)
\label{eq2}
\end{equation} 
It is clear that Eq. \ref{eq2} is invariant under general rotations in the $(\sqrt{\phi(x)},v)$ space. Let's now combine this rotational symmetry
in the $(\sqrt{\phi(x)},v)$ space with the cold symmetry in the initial conditions. In the cold initial conditions the density is a delta function
in the velocity space. The delta function will be un-changed by a re-scaling of the velocity axis, the only modification due to the re-scaling
of the velocity space will be a scaling of the amplitude of the function. As a consequence a re-scaling of the velocity will produce the same final
density at equilibrium up to a scaling factor. Now if we include the rotational symmetry at equilibrium, the phase-space density must also
be invariant by general rotations in the $\left(\sqrt{\phi(x)},v \right)$ space. Let first make a re-scaling in velocity,
\begin{equation}
 v \rightarrow \tilde v=\alpha  v \\
 \label{eq3.0}
\end{equation}
We also consider the effect of the velocity re-scaling on the potential $\phi(x)$,
\begin{equation}
 \phi(x) \rightarrow \tilde \phi(x)\\
 \label{eq3.00}
\end{equation}
By applying the velocity transformation in Eq. (\ref{eq3.0}) to Eq (\ref{eq2}) we obtain,
\begin{equation}
 f(\tilde \phi(x)+ \alpha^2 v^2)= \beta f(\phi(x)+v^2) = \beta f(E)
 \label{eq3.1}
\end{equation}
Let now consider the special case $\phi(x)=0$, then Eq. (\ref{eq3.1}) reads,
\begin{equation}
 f(\alpha^2 v^2)= \beta f(v^2)
 \label{eq4}
\end{equation}
Note that we assumed implicitly that $\phi=0$ implies that $\tilde \phi=0$. In 1 dimension we consider initial conditions with symmetry with respect to the origin. In radial coordinates in muti-dimension the same symmetry is implicit. The value $\phi=0$ is observed at the center of the system and is a consequence of the symmetry of the system.  The velocity re-scaling that we apply conserve this symmetry thus $\tilde \phi=0$.
Let now apply a rotation in the $\left(\sqrt{\phi(x)},v \right)$ space. Then $(\phi=0 ,v)$ is transformed to $(\sqrt{\phi_1(x)}, v_1)$, the norm
is conserved in the rotation thus, 
\begin{equation}
 \phi_1+v_1^2=v^2. 
 \label{eq5}
\end{equation}
By introducing Eq. (\ref{eq5}) in Eq. (\ref{eq4}) we obtain,
\begin{equation}
 f\left(\alpha^2 \left( \phi_1(x)+v_1^2\right) \right)= \beta f\left(\phi_1(x)+v_1^2\right)
 \label{eq6}
\end{equation}
Which leads directly to:
\begin{equation}
 f\left(\alpha^2  E_1 \right)= \beta f\left(E_1\right)
 \label{eq7}
\end{equation}
It is clear that (\ref{eq7}) express the full self-similarity of the phase space density at equilibrium.
Note that the transformation of $\phi$ due to the velocity re-scaling can be obtained using Eq. (\ref{eq6}).
The left side of the equation express the phase space density in the transformed coordinates $(\tilde \phi, \tilde v)$
By direct identification we find the transformed potential $\tilde \phi$ (see Eq. \ref{eq3.00}),
\begin{equation}
 \tilde \phi(x) = \alpha^2 \phi(x)
 \label{eq8}
\end{equation}
An equivalent transformation is to consider the effect of the velocity re-scaling on the variable x. Let's define
$\tilde x$ as the transformation of x. Then,
\begin{equation}
 \tilde \phi(x) =\phi\left(\tilde x \right)= \alpha^2 \phi(x)
 \label{eq9}
\end{equation}
Let's now consider the Poisson equation for a n dimensional system,
\begin{equation}
 x^{1-n} \frac{d}{d x} \left(x^{n-1} \frac{d \phi}{d x} \right)=\rho=\int f d^n v
\label{eq10}
\end{equation}
Let then apply the velocity re-scaling transformation to Eq. (\ref{eq10}),
\begin{equation}
 \tilde x^{1-n} \frac{d}{d \tilde x} \left(\tilde x^{n-1} \frac{d \tilde \phi}{d {\tilde x}} \right)=\tilde \rho=\int \tilde f d^n \tilde v
\label{eq11}
\end{equation}
Using Eq's (\ref{eq9}), (\ref{eq7}) and (\ref{eq10}) we obtain,
\begin{equation}
\alpha^2 \left(\left(\frac{d x}{d \tilde x}\right)^2 \frac{d^2 \phi}{d x^2} +{\tilde x}^{1-n} \frac{d}{d \tilde x}\left( {\tilde x}^{n-1} \frac{d x}{d \tilde x} \right) \frac{d \phi}{d x} \right) =\beta \alpha^n x^{1-n} \frac{d}{d x} \left(x^{n-1}\frac{d \phi}{d x} \right)
\label{eq12}
\end{equation}
Eq. (\ref{eq12}) has two type of solutions for the transformation $\tilde x$, (I) solutions of the type $\frac{d x}{d \tilde x}={\rm Constant}$, or (II) general solutions which depends on $\phi$. As an illustration for $n=1$ the type (II) general solution reads:
\begin{equation}
\frac{d x}{d \tilde x} = \beta \alpha^{n-2} \tanh \left[\log\left( \frac{d \phi}{d x}\right) +C_1\right]
\label{eq13}
\end{equation}
The problem with the solution of type (II) in Eq. (\ref{eq13}) is that for a general potential $\phi(x)$ the identity condition
(in the abscence of re-scaling) $\tilde x(x)=x$ for $\beta=1$ and $\alpha=1$ cannot be universally satisfied. Thus we are left
with only the type (I) solution where $\tilde x$ is a linear function of $x$. This type (I) solution reads, 
\begin{equation}
\tilde x = {\beta}^{-\frac{1}{2}} \alpha^{-\frac{n}{2}+1} x=\gamma x
\label{eq14}
\end{equation}
Eq. (\ref{eq14}) implies the similarity of the potential, using Eq. (\ref{eq9}) we find,
\begin{equation}
\phi\left(\gamma x \right)= \alpha^2 \phi(x)
\label{eq15}
\end{equation}
Similarly the similarity of $f$ may be re-expressed,
\begin{equation}
f\left(\gamma x, \alpha v\right) = \beta f\left(x,v\right)
\label{eq16}
\end{equation}
\subsection{System depending on other integrals.}
 If the phase space density depends on $E$ and the angular momentum $J$ Eq. (\ref{eq2}) now reads,
\begin{equation}
 f(x,v) \equiv F(E,J) \equiv F \left(\phi(x)+v^2, |\bf r \times v| \right)
\label{eq17}
\end{equation}
Similarly Eq. (\ref{eq3.1}) now reads,
\begin{equation}
f\left(\tilde \phi(x)+ \alpha^2 v^2,\tilde J\right)= \beta f(\phi(x)+v^2,J) = \beta f(E,J)
\label{eq3.1a}
\end{equation}
Starting from here we will apply the reasonning we already used in Sec. (\ref{sec_1}). We note the similarity between Eq. (\ref{eq3.1}) and Eq. (\ref{eq3.1a}) and proceed through equations (\ref{eq4}) (\ref{eq5}) (\ref{eq6}) and (\ref{eq7}). By this process we arrive at a first similarity equation for $f$,
\begin{equation}
f\left(\alpha^2 E,\tilde J\right) = \beta f(E,J)
\label{eq18}
\end{equation}
The Poisson equation is not modified with respect to Sec. (\ref{sec_1}) resulting in the same results for the similarity
of the potential and of the $x$ variable (Eq's \ref{eq15} and \ref{eq16}).
Using Eq's (\ref{eq16}) and Eq. (\ref{eq3.0}) we find that angular momentum re-scacle like:
\begin{equation}
\tilde J=\alpha \gamma J
\label{eq19}
\end{equation}
By combining Eq. (\ref{eq18}) and Eq. (\ref{eq19}) we arrive at a full similarity for $f$,
\begin{equation}
f\left(\alpha^2 E, \alpha \gamma J\right) = \beta f(E,J)
\label{eq20}
\end{equation}
\section{Convergence towards auto-similarity.}
\label{Sec_Conv}
\begin{figure*}
 \includegraphics[width=17cm]{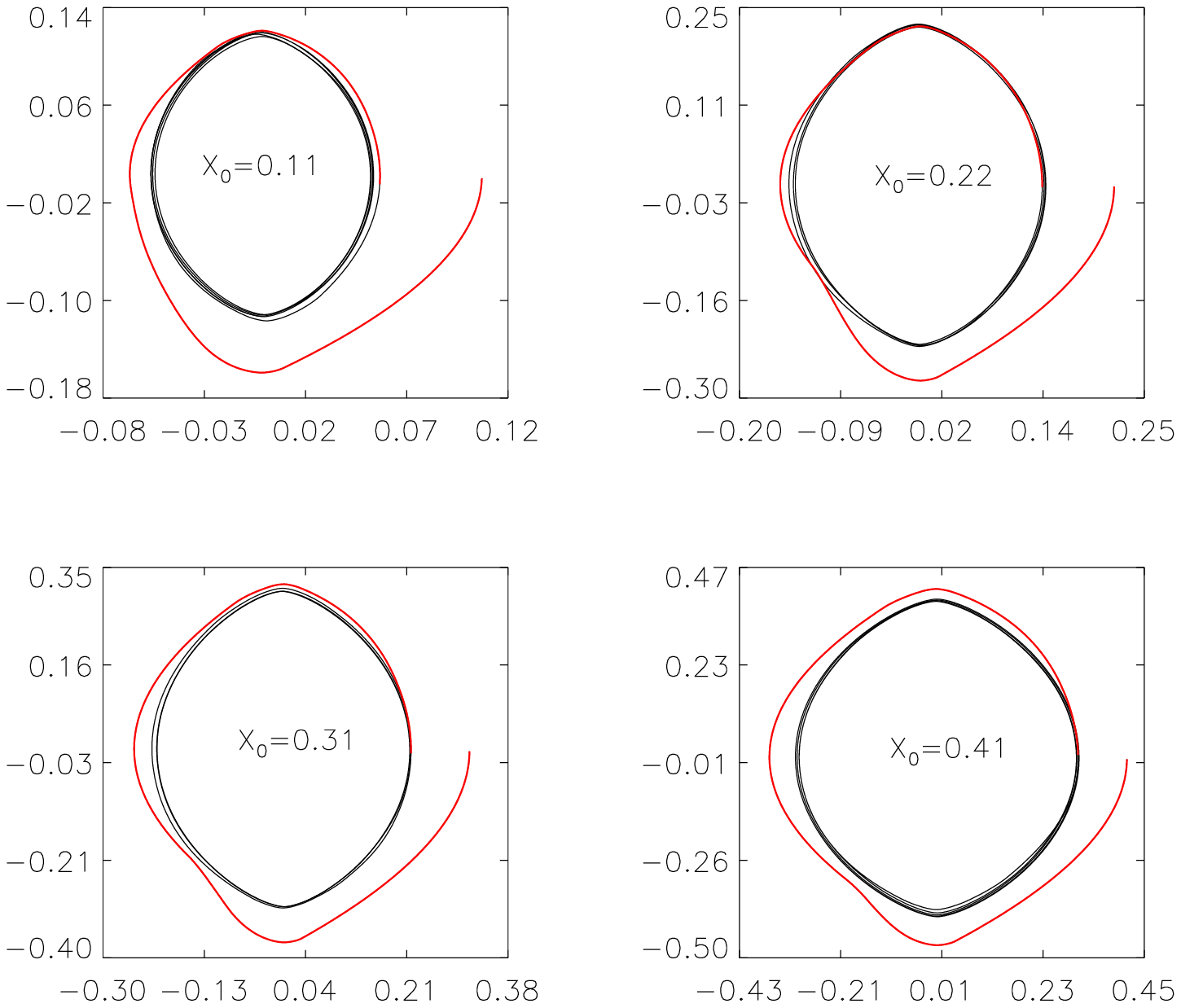}
 \caption{The trajectory in the $(x,v)$ phase space for points within the central region.
 The initial density is a power-law profile with exponent $-0.35$.
 The presented trajectories start from the initial conditions, the first orbit is in red color. The value of $X_0$ in each graph indicates the starting position in the initial conditions. In all cases the stationary regime is reached approximately after one orbital time.}
\label{Fig3}
\end{figure*}

The ideas and calculations developed in Sec. (\ref{Sec_Eq}) indicates that self-similarity is a consequence of equilibrium for cold initial conditions. However self-similar solutions belong to a large class of solutions and the process leading to the convergence towards a specific solution
is not clear. Here we propose a numerical study of this process for 1D gravity. The simulations are identical to the 1D simulations produced in \cite{Alard2013a} (see Sec. 9). A fundamental problem is to asses the number of orbital times required to reach equilibrium. Since we already know
that equilibrium is equivalent to self-similarity it is also a way to asses the number of orbital time required to reach self-similarity. A general result obtained from the 1D simulations is that the convergence towards a stationary regime is obtained within only one orbital time (see Fig \ref{Fig3}). Since the orbital time is a decreasing function of the distance to the center, the time required for the convergence towards a stationary regime become infinitely small towards the center. The energy of the particles on orbits very close
to the center is also very small. This is indeed the case in 1D if the density is a power-law with exponent less than $-2$. This
is also the self-similar solution in 3D observed in numerical simulations. This solution has an asymptotic power-law slope in density of $-1$.
As a consequence the solution converges first in an area infinitely close to the center in an infinitely short time and for infinitely small
energies. Once a solution has developed in this area very close to the center, the degeneracy on the choice of the self-similar
solution is broken. The breaking of the degeneracy itself requires very little energy since it involves modifying the dynamics of particles with infinitely small energies. As a consequence the nature of any perturbating process near the center of the system
is very important and will drive the convergence towards a specific self-similar solution. An example of a process acting in the central region and leading to a general convergence towards an universal solution is illustrated in \cite{Alard2013a}.
In this case if the density in the initial conditions are less concentrated than a power law of exponent $-\frac{1}{2}$, a general convergence towards a power-law density with exponent $-\frac{1}{2}$ is observed at equilibrium. The process
acting in the central region and leading to the convergence is due to the density profile of caustics very close to the center. Another example is the convergence towards an universal profile for a spherically symmetrical system with constant
angular momentum. This time the convergence is due to the dominating effect of the angluar momentum in the central region (see \cite{Halle}).
\section{Conclusion.}
This work offers an explanation for the self-similar conjecture observed for cold initial conditions. Even if the density of these initial
conditions is not self-similar, which would require a power-law density, the system evolves towards self-similarity. The explanation
is that the initial cold-symmetry of the initial conditions when combined with the symmetry of the equilibrium state induces a new class
of symmetry which is equivalent to self-similarity. This result is general it was demonstrated in Sec. \ref{Sec_Eq} that self-similarity
occur at equilibrium for cold initial conditions in 1D or spherically symmetric gravity in 3D. On the other hand if self-similarity is predicted
at equilibrium it does not specify the self-similar solution itself. The self-similar solutions belong to a large class of solutions and the convergence towards a specific solution must be induced by some process. An illustration of this convergence process in 1D gravity is provided in Sec. \ref{Sec_Conv}. The convergence of the solution is operated in a typical orbital time. Since the orbital
time become very small at the center the breaking of the degeneracy and the convergence towards a specific solution is first operated
in the central region, and then propagates towards the outer regions of the system. The initial change of the orbit at the center requires
only an asymptotically small change in energy.
\section*{Data Availability}
No datasets were generated or analyzed during the current study.
\end{document}